\documentclass[aps,twocolumn,showpacs,showkeys]{revtex4}
\usepackage{graphicx}

\begin{document}

\title{Scaling of the dynamical properties of the Fermi-Ulam accelerator model}

\author{Denis Gouv\^ea Ladeira$^{1,\ast}$ and Jafferson Kamphorst Leal da Silva$^{1,\dagger}$}

\affiliation{
$^{1}$ Departamento de F\'\i sica, ICEx\\
Universidade Federal de Minas Gerais\\
Caixa Postal 702, 30.123-970, Belo Horizonte/MG, Brazil\\}

\date{\today}\widetext



\begin{abstract}

The chaotic sea below the lowest energy spanning curve of the complete Fermi-Ulam model 
(FUM) is numerically investigated when the amplitude of oscillation $\varepsilon$ of
the moving wall is small. 
We use scaling analysis near the integrable to non-integrable transition to describe 
the average energy as function of time $t$ and as function of iteration (or collision) number $n$. 
If $t$ is employed  as independent variable, the exponents related 
to the  energy scaling properties of the FUM are different from
the ones of  a well known simplification of this model (SFUM).
However, if $n$ is employed as independent variable, the exponents are the same for both FUM and SFUM.
In the collision number analysis, we  present analytical arguments supporting that the 
exponents $c^*$ and $b^*$ related to the initial velocity and to $\varepsilon$
are given by $c^*=-1/2$ and $b^*=-1$.  We derive also a relation connecting the scaling 
exponents related to the variables time and collision number. 
Moreover, we show that, differently from the SFUM, the average energy in 
the FUM {\sl saturates} for long times and we justify 
the physical origins for some differences and similarities observed between the 
FUM and its simplification. 

\end{abstract}

\maketitle

\section{Introduction}

The Fermi accelerator is a dynamical system, proposed by 
Enrico Fermi \cite{fermi} to describe cosmic ray acceleration, in 
which a charged particle collides with a time dependent 
magnetic field. With base in different applications, the 
original model was later modified and studied by other 
authors. One of them is the well known Fermi-Ulam model (FUM) \cite{ulam1,ulam2}, 
in which a ball is confined between a rigid fixed wall and an 
oscillatory moving one.
In this model the ball collides elastically with the walls 
and the system is described by an area preserving map.
In high energies regime the phase space presents 
Kolmogorov-Arnold-Moser (KAM) islands 
surrounded by locally chaotic regions, which are limited 
by spanning curves. Below the invariant spanning curve of 
lower energy a globally chaotic sea involves KAM islands.
The presence of invariant spanning curves limits the orbits 
in phase space impeding unlimited energy growth (Fermi 
acceleration) \cite{ulam2,douady}. 
Another model, very similar to FUM, in which the ball is 
in a gravitational field, so-called bouncer \cite{bouncer1}, presents, 
differently from FUM, the property of Fermi acceleration 
depending on values of control parameter and initial 
conditions. 
This difference between the two models was later explained 
by Lichtenberg et al \cite{bouncer2}. 
Hybrid versions \cite{leonel3} involving both FUM and bouncer and versions 
of FUM with energy dissipation \cite{leonel4} were also explored. 
Quantum models based on FUM and bouncer have also been 
studied \cite{quantum1,quantum2,quantum3}.
The study of such systems is interesting because it allows 
to compare theoretical predictions with experimental results \cite{exp1,exp2}; 
moreover the knowledge about how time-dependent perturbations 
affect the dynamics of Hamiltonian systems is something that 
needs to be more explored. 
Therefore it is useful to study such perturbations in simple 
models because they furnish insights about more 
complex systems; even more, the formalism used in 
its characterization can be extended to the billiard class 
of problems \cite{bilhar1,bilhar2,bilhar3,bilhar4}. 

The simplified Fermi-Ulam model (SFUM) \cite{ulam1} is an approximation in which 
the position of the \textit{moving wall} 
is considered as \textit{fixed}, but it transfers momentum and energy to the particle.
This geometrical change of the complete model can be neglected 
when the amplitude of oscillation is much smaller than the distance between the two 
walls and the velocity of the particle is larger than the wall maximal velocity.
It is clear that the simplified versions can speed up the simulations. However the 
main interest in these simplified models is that they can be studied 
by analytical methods whose results are often compared with the 
numerical results of the simplified model. Moreover, these analytical results 
are, sometimes, also useful in study of the full models. 

The moving wall in the Fermi-Ulam model represents an external force 
that acts as a perturbation in the system. 
If the oscillation amplitude of the moving wall is zero 
the system is integrable but as soon as this amplitude 
is different from zero the Fermi-Ulam model behaves chaotically \cite{leonel1}. 
Near the integrable to non-integrable transition, average quantities 
can be described by scaling functions.
This kind of analysis was originally proposed in a 
study of the SFUM, where the collision number with the moving 
wall is employed as independent variable \cite{leonel2}. 
In a recent work \cite{ladeira}, also based on the SFUM, we proposed 
a similar analysis in which time is considered as 
independent variable and we showed 
that the average energy can also be described by scaling functions 
but with a different exponents set. 
Moreover, we showed that the average energy 
{\sl decays} at long times. 

In the present work we investigate the chaotic sea below 
the spanning curve of lowest energy for the full Fermi-Ulam model, 
where scaling properties of average energies are studied on variables 
\textsl{time} and \textsl{iteration} (or \textsl{collision}) \textsl{number}. 
We show that if the time is employed as independent variable, then 
the exponents related to the 
scaling properties of FUM are not the same that the ones of SFUM. 
However, if the iteration number is employed as independent variable, 
then both FUM and SFUM present the same set of exponents. 
We show that the energy decay found for long time in the SFUM \cite{ladeira} 
does not exist for the FUM and 
we justify physically the origins of the similarities and 
differences between the FUM and SFUM. 
We provide also some analytical results for the scaling exponents 
and, moreover, we show that although the exponents related to the variables 
\textsl{time} and \textsl{collision number} 
are not the same a relation between them can be established. 

This paper is organized as follows: In the next section the Fermi-Ulam 
model and the average quantities of interest are defined. We present also the procedures 
to obtain these averages when time is the independent variable. 
In Section III we present the results of the numerical simulations, 
the scaling properties of average energies as function of both 
time and collision number, and we determine the exponents related 
to these averages. 
In Section IV analytical arguments to determine some exponents 
and to derive a relation between the exponents of time and 
collision number analyses are discussed. 
Finally, we draw the conclusions in Section V, 
where we also present a summary of this present work. 

\section{The Fermi-Ulam model}

The Fermi-Ulam model describes the motion of 
a classical particle bouncing between two rigid walls, 
one of which is fixed at position $x=0$ and other 
that is moving periodically in time whose position 
is given by 
$x_w(t^{\prime\prime})=x_0 +\varepsilon^\prime\cos(\omega 
t^{\prime\prime}+\phi_0)$. 
Here $x_0$ is the equilibrium position, 
$\varepsilon^\prime$ is the oscillation amplitude, 
$\omega$ is a frequency, $t^{\prime\prime}$ is time 
and $\phi_0$ is the initial phase. 
In order to work with dimensionless variables we 
perform scale changes in length $X_w=x_w/x_0$ and 
in time $t^\prime=\omega t^{\prime\prime}$. 
With these new variables the system has just one 
parameter, namely $\varepsilon=\varepsilon^\prime/x_0$, 
and we write the position of the moving wall as 
$X_w(t^\prime)=1+\varepsilon\cos(t^\prime+\phi_0)$.
The particle moves freely between the walls and 
collides elastically with them. 
In this manner the FUM can be described by a map 
$T(V_n,\phi_n)=(V_{n+1},\phi_{n+1})$ which gives 
the velocity of the particle and the phase of the 
moving wall immediately after each collision \cite{leonel1} 
\begin{equation}
T=\left\{\begin{array}{ll}
V_{n+1}=\pm V_n+2\varepsilon\sin(\phi_{n+1})~~\\
\phi_{n+1}=\phi_n+\Delta t_{n+1}~~~ \rm{mod}~2\pi\\
\end{array}
\right..
\label{eq1}
\end{equation}

Here the term $2\varepsilon\sin(\phi_{n+1})$ 
gives the fraction of velocity gained or lost 
in collision and $\Delta t_{n+1}=t_{n+1}-t_{n}$ is 
the time between two collisions with the moving 
wall, which is given by the smallest solution of 
\begin{equation}
V_n\Delta t_{n+1}-(1+\varepsilon\cos\phi_n)=\pm [1+\varepsilon\cos(\Delta t_{n+1}+\phi_n)].
\label{eq2}
\end{equation}

The plus sign in the above equations  corresponds to the situation in which the particle collides 
with the fixed wall before hitting the moving one (\textit{indirect collisions}); 
the minus sign corresponds to the situation in which the particle hits successively with the 
moving wall (\textit{direct collisions} or \textit{successive collisions}). 

Let us define $V^2(t^\prime)$ as the square velocity at time $t^\prime$. 
We are interested in the scaling properties of the dimensionless energy 
$E=2~\rm{Energy}/m\omega^2 x_0^2$ averaged 
over an ensemble of $M$ samples that belong to the chaotic sea. 
Such samples are characterized by initial 
phases $\phi_0$ of the moving wall randomly 
chosen in an interval $I$. 
If the initial velocity $V_0$ is small enough then 
we can use $I=[0,2\pi)$. 
Moreover we consider that the particle starts at the 
fixed wall position with velocity $V_0>0$. 
Thus, the time of the first collision of the 
particle with the moving wall is given by 
$T_1\approx 1/V_0$ and we define a variable $t$ as 
$t=t^\prime-1/V_0$. 
In this way the time $t$ starts at the first collision 
instant and we write the average energy as
\begin{equation}
E(t,\varepsilon,V_0)={1\over M}\sum_{j=1}^M V^2_j(t)~~,
\label{eq3}
\end{equation}
where $j$ refers to a sample. 

We consider also another kind of average where the 
square velocity is firstly averaged over the orbit of a 
sample as 
\begin{equation}
\overline{V^2}(t^\prime)={1\over{t^\prime}}\int_0^{t^\prime} V^2(\tau )~d\tau~~,
\label{eq4}
\end{equation}
and then we perform the average of the  energy
in an ensemble of $M$ samples as defined below 
\begin{equation}
\overline{E}(t,\varepsilon,V_0)={1\over M}\sum_{j=1}^M\overline{V_j^2}(t)~~.
\label{eq5}
\end{equation}

As the particle moves freely between the walls, the square of its 
velocity is constant between two impacts with the moving wall and 
the integral in Eq. (\ref{eq4}) can be numerically evaluated without difficulties. 

We will now describe some numerical procedures 
used to evaluate the averages quantities on time. 
The dynamics of FUM as described in Eq. (\ref{eq1}) evolves in a 
discrete variable, namely the collision number $n$.
As we are interested in the evolution of the average energies 
as defined in Eqs. (\ref{eq3}) and (\ref{eq5}), where time $t$ is 
a continuous variable, we can use the map given by Eq. (\ref{eq1}) 
to speed up the calculation process. 
We evaluate the averages in Eqs. (\ref{eq3}) and (\ref{eq5}) at 
discrete, logarithmic spaced, values of time 
$t=t_1,~t_2,~\ldots,~t_N$. 
We known that at $t=0$, the first collision instant, 
the energy of the particle is $V_1^2$ and, as the particle is at 
the moving wall position, the next collision 
takes place at $t_c=\Delta t_2$, where $\Delta t_2$ is obtained by solving Eq. (\ref{eq2}). 
To evaluate the energy $E_1$ of the first sample of the ensemble 
at time $t=t_1$ we follow the procedure: 
(1) If $t_1\leq t_c$, the energy of the particle at $t_1$ 
        will be $E_1=V_1^2$. 
(2) Otherwise $t_1 > t_c$ and a collision occurs at time $t=t_c=\Delta t_2$. 
        Then we employ Eq. (\ref{eq1}) updating the velocity of the 
        particle to $V_2$ and the 
        next collision time is given by $t_c = \Delta t_2 + \Delta t_3$. 

It means that if case 1 is satisfied then $E_1(t_1)$ was determined. 
Otherwise we update the next collision instant and the velocity of the 
particle. 
If now $t_1\leq t_c$ (case 1) then $E_1=V_2^2$ but if $t_1$ is still larger than $t_c$, 
we repeat case 2 updating both $V$ and $t_c$. The reasoning is 
basically to repeat the above procedure until $t_1\leq t_c$ and then 
to update the energy $E_1(t_1)$. 
We follow a similar proceeding, with appropriate time intervals, 
to evaluate $E_1(t_2),~E_1(t_3),~\ldots,~E_1(t_N)$. 

The procedure is the same for all samples of the ensemble giving 
$E_2,~\ldots,~E_M$ from $t_1$ until $t_N$. Then the average energy 
in Eq. (\ref{eq3}) is performed doing 
$E(t_1)=[E_1(t_1)+E_2(t_1)+\ldots+E_M(t_1)]/M,~\ldots,~
 E(t_N)=[E_1(t_N)+E_2(t_N)+\ldots+E_M(t_N)]/M$. 
The average energy $\overline{E}(t,\varepsilon,V_0)$ in Eq. (\ref{eq5}) 
can be evaluated by a similar procedure.

Since the map in Eq. (\ref{eq1}) gives the velocity and phase after each collision 
with the moving wall, then the averages on variable $n$ can be defined in a more 
direct manner. 
We first consider the average velocity over the orbit generated from an initial phase 
$\phi_0$, defined as $\overline{V^2}(n)={1\over n+1}\sum_{i=0}^{n} V_i^2$. 
Considering an ensemble with $M$ samples, characterized by initial conditions that 
belong to the chaotic sea, we define the average energy as 
\begin{equation}
<E>(n,\varepsilon,V_0)={1\over M}\sum_{j=1}^M\overline{V_j^2}(n)~~.
\label{eq9}
\end{equation}

\section{Results}

\begin{figure}[t]
\centerline{\includegraphics[width=7cm,height=4.5cm]{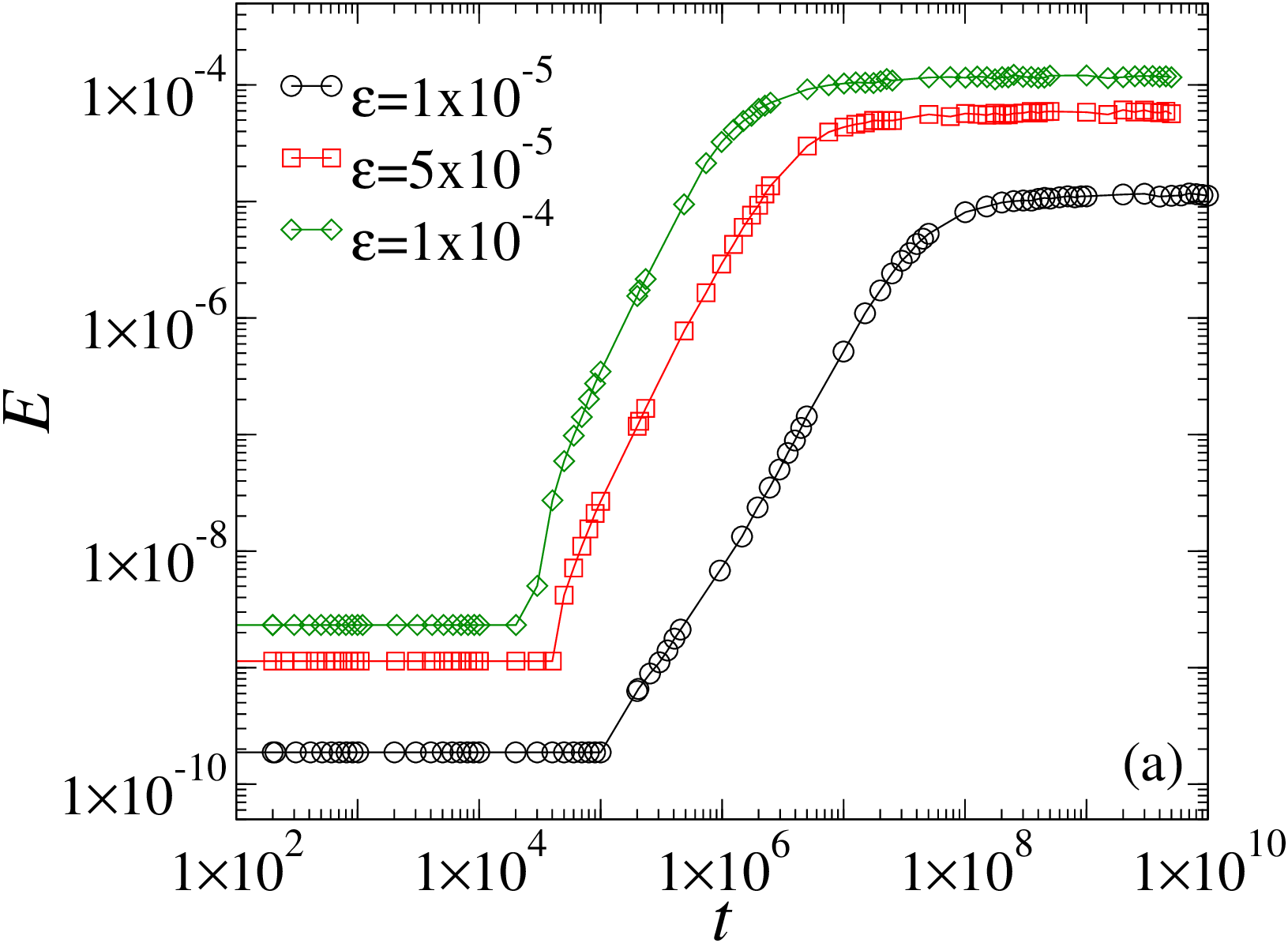}}
\vspace{0.4cm}
\centerline{\includegraphics[width=7cm,height=4.5cm]{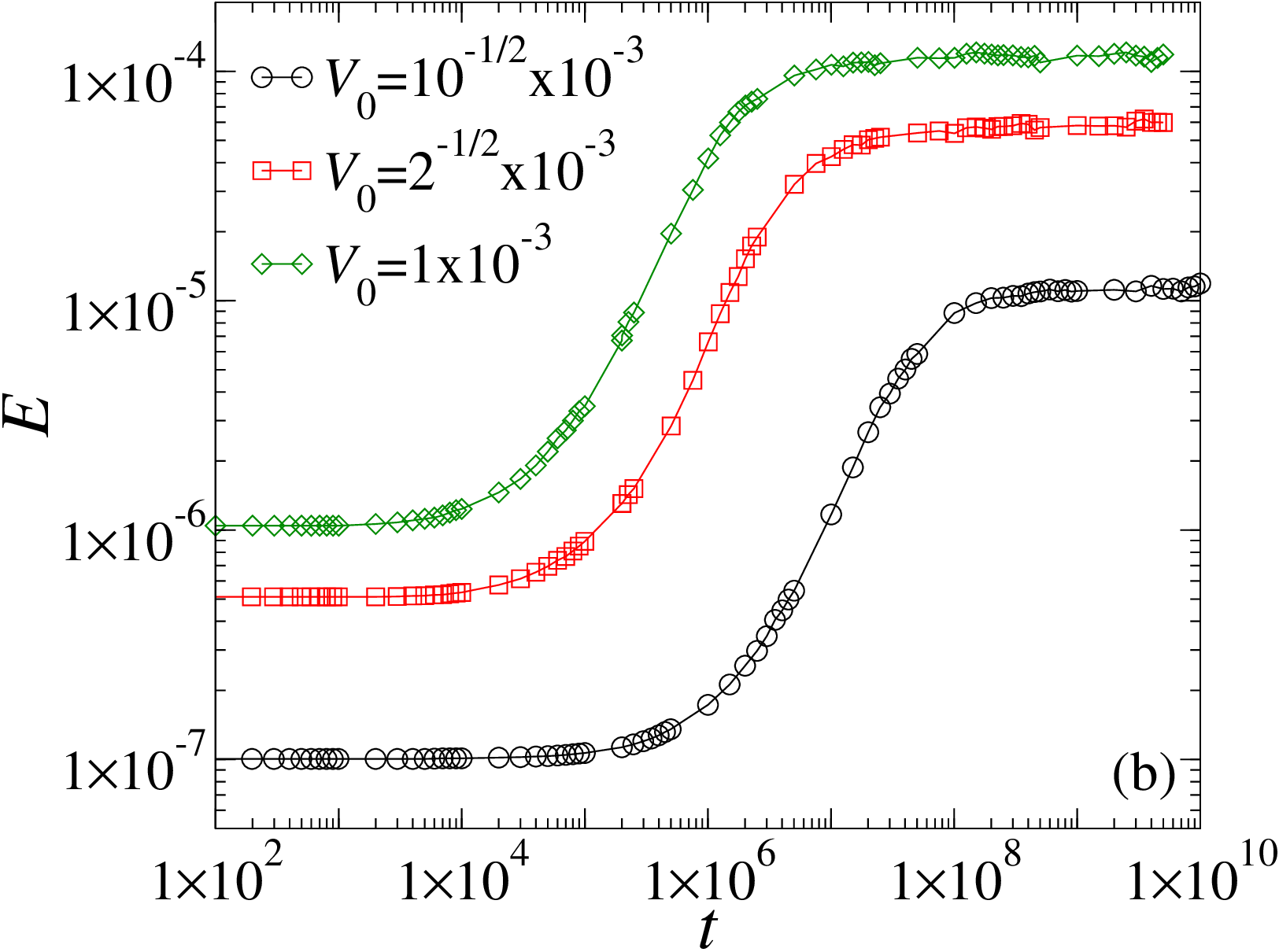}}
\caption{(Color online) Log-log plots of the energy $E(t)$ defined in 
Eq. (\ref{eq3}), averaged over $M=2\times 10^3$ samples, as function of time 
for (a) three values of $\varepsilon$ and $V_0=1\times 10^{-6}$ 
($V_0<<\varepsilon$) and (b) three values of $\varepsilon$, the same ones 
as those shown in (a), and three values of initial velocity 
$V_0>>\varepsilon$.}
\label{fig1}
\end{figure}

In Fig. \ref{fig1}(a) and \ref{fig1}(b) we show, respectively, 
the numerical results for the average energy 
$E(t,\varepsilon,V_0)$ when the initial velocity is small, or 
$V_0<<\varepsilon$, and the situation in which $V_0>>\varepsilon$. 
The averages defined in Eqs. (\ref{eq3}) and (\ref{eq5}) were 
performed in an ensemble with $M=2\times 10^3$ samples. 
Although $V_0>>\varepsilon$, the energy curves shown in 
\ref{fig1}(b) are obtained from orbits that belong to the chaotic sea. 
In Fig. \ref{fig1}(b) we used the same values of $\varepsilon$ as 
those shown in Fig. \ref{fig1}(a). 

As we can see more clearly in Fig. \ref{fig1}(a) the energy is 
constant until a time $t_1$, grows up to a time $t_2$ and then
reaches a stationary value for large values of time. 
As show in \cite{ladeira} the energy in the simplified FUM presents a 
slow decay in time. 
Since in SFUM the 
oscillating wall is considered fixed at position $x=1$, the 
time between two collisions with the oscillating wall is 
$2/V_n$, and successive collisions do not occur. 
Therefore, if after a collision the particle has very low velocity, 
then it remains for a long time, $2/V_n$, with low energy. 
In this way, at time $t$ many realizations are in such condition originating 
the slow decay in the average energy for $t>>t_2$. 
In FUM this situation does not occur. If the particle, 
after a collision, losses almost all its energy, 
a successive collision occurs increasing the energy of the particle. 
Therefore, in FUM the average energy does not decay, but presents a 
saturation regime for $t>>t_2$. 
Moreover, the energy for $t>>t_2$ can be described as 
\begin{eqnarray}
 E(t,\varepsilon,V_0)&\propto& g(\varepsilon)~~,\nonumber\\
 g(\varepsilon)&\propto&\varepsilon^\beta~~.
 \label{eq6}
\end{eqnarray}

\begin{figure}[t]
\centerline{\includegraphics[width=7cm,height=4.5cm]{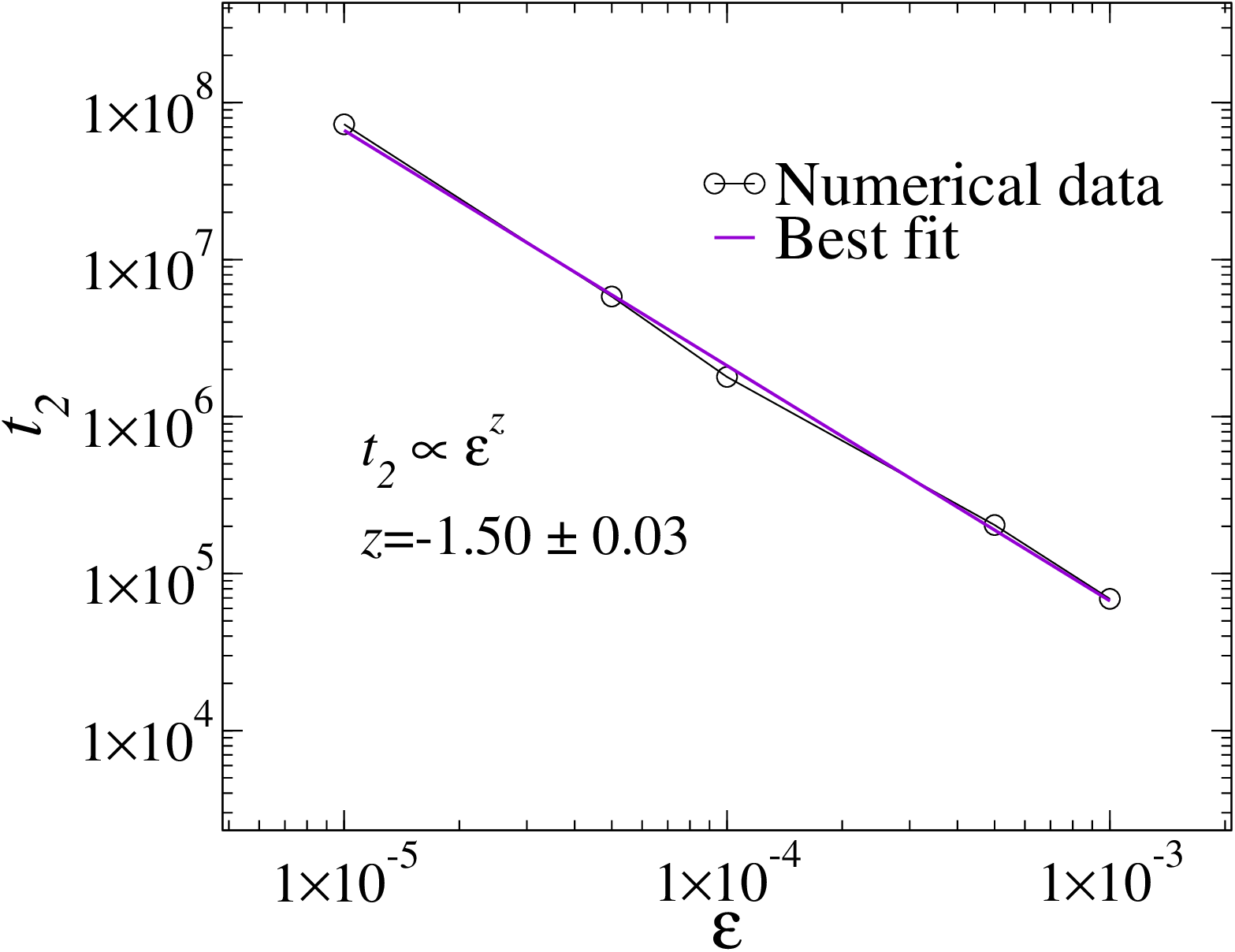}}
\caption{Log-log plot of the crossover time $t_2$ as function of 
parameter $\varepsilon$. }
\label{te}
\end{figure}

The value of exponent $\beta$ can be determined by searching 
for the best collapse of the energy curves in the asymptotic regime. 
We performed simulations for values of $\varepsilon$ 
between $1 \times 10^{-5}$ and $1 \times 10^{-3}$ 
and we obtained the average value $\beta=1.01\pm 0.01$. 
The crossover time $t_2$ obeys the relation 

\begin{equation}
t_2\propto\varepsilon^{z}~~,
\label{eq7}
\end{equation}
and the better fit in a plot of $t_2$ as a function of $\varepsilon$ 
gives $z=-1.50\pm 0.03$, as shown in Fig. \ref{te}. 

The time $t_1$ is an average time in which the second indirect collisions 
happen. After that new collisions occur and the energy $E(t,\varepsilon,V_0)$ 
begins to increase. Moreover, $t_1$ is related to an initial transient and 
it affects the behavior of $E(t,\varepsilon,V_0)$ for $t<<t_2$. Therefore, 
between $t_1$ and $t_2$ we have a crossover region in which the growth exponent 
can not be directly evaluated. 

For large values of time ($t>>t_1$), the energy $E(t,\varepsilon,V_0)$ 
can be written as a scaling function, namely 
\begin{equation}
E(t,\varepsilon,V_0)= lE(l^{a}t,l^{b}\varepsilon,l^{c}V_0)~~.
\label{eq8}
\end{equation}
Here $a$, $b$ and $c$ are scaling exponents and $l$ is the scaling factor. 
In the limit $V_0<<\varepsilon$ (Fig. \ref{fig1}(a)) we can choose 
$l=\varepsilon^{-1/b}$ and write the above equation as 
$E(t,\varepsilon,0)= \varepsilon^{-1/b}E(\varepsilon^{-a/b}t,1,0)\propto 
\varepsilon^{-1/b}f(\varepsilon^{-a/b}t)$. 
For $t>>t_2$ this relation can be written as 
$E(t,\varepsilon,0)\propto\varepsilon^{-1/b}$. 
Then, from Eq. (\ref{eq6}) we have that $\beta=-1/b$. Using the simulation 
value $\beta=1.01\pm 0.01$ we obtain $b=-0.99\pm 0.01$. 
From Eq. (\ref{eq7}) we derive the relation $z=a/b$. Since the 
simulations furnish $z=-1.50\pm 0.03$, it follows that $a=1.49\pm 0.04$. 

\begin{figure}[t]
\centerline{\includegraphics[width=7cm,height=4.5cm]{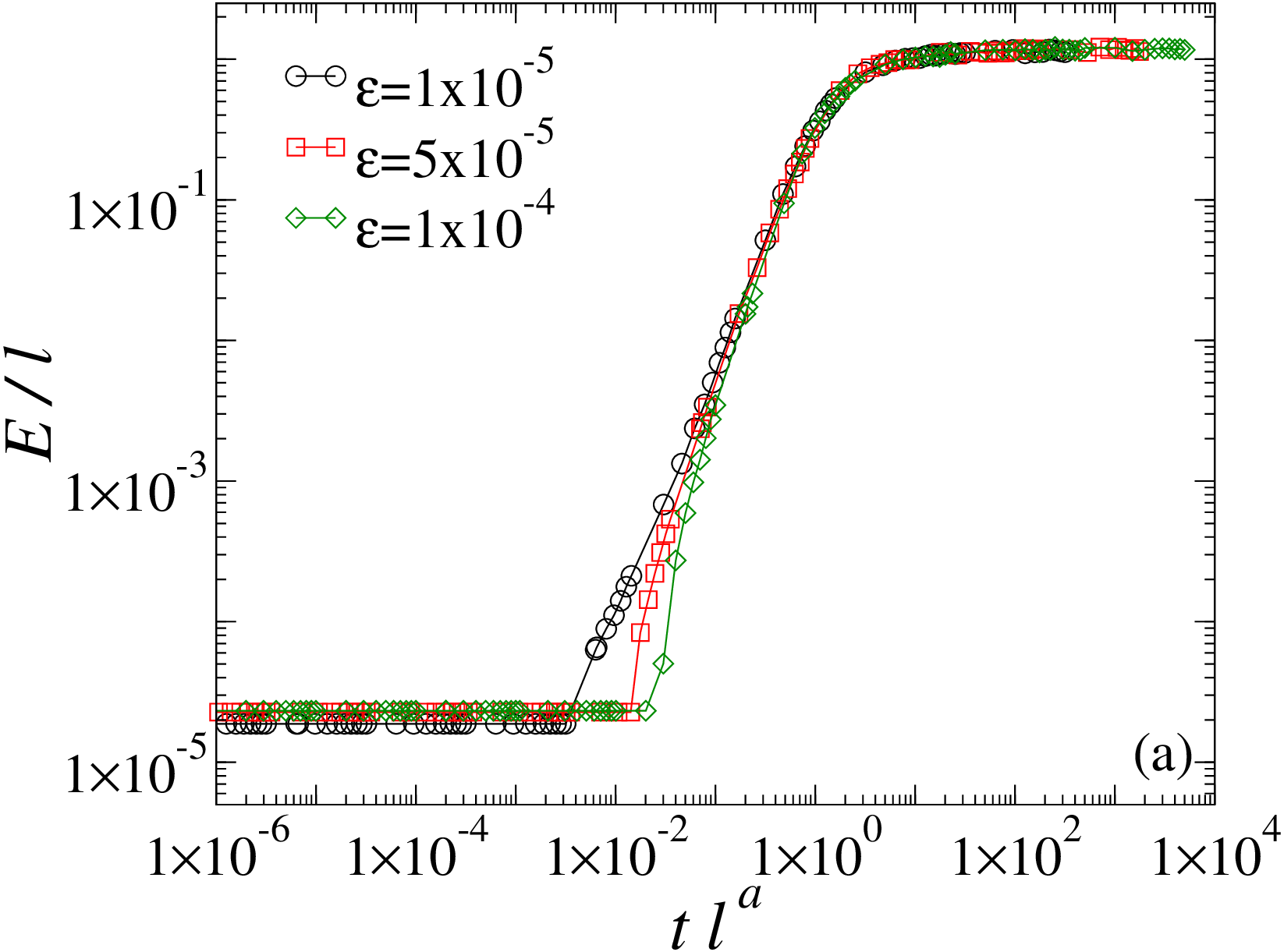}}
\vspace{0.4cm}
\centerline{\includegraphics[width=7cm,height=4.5cm]{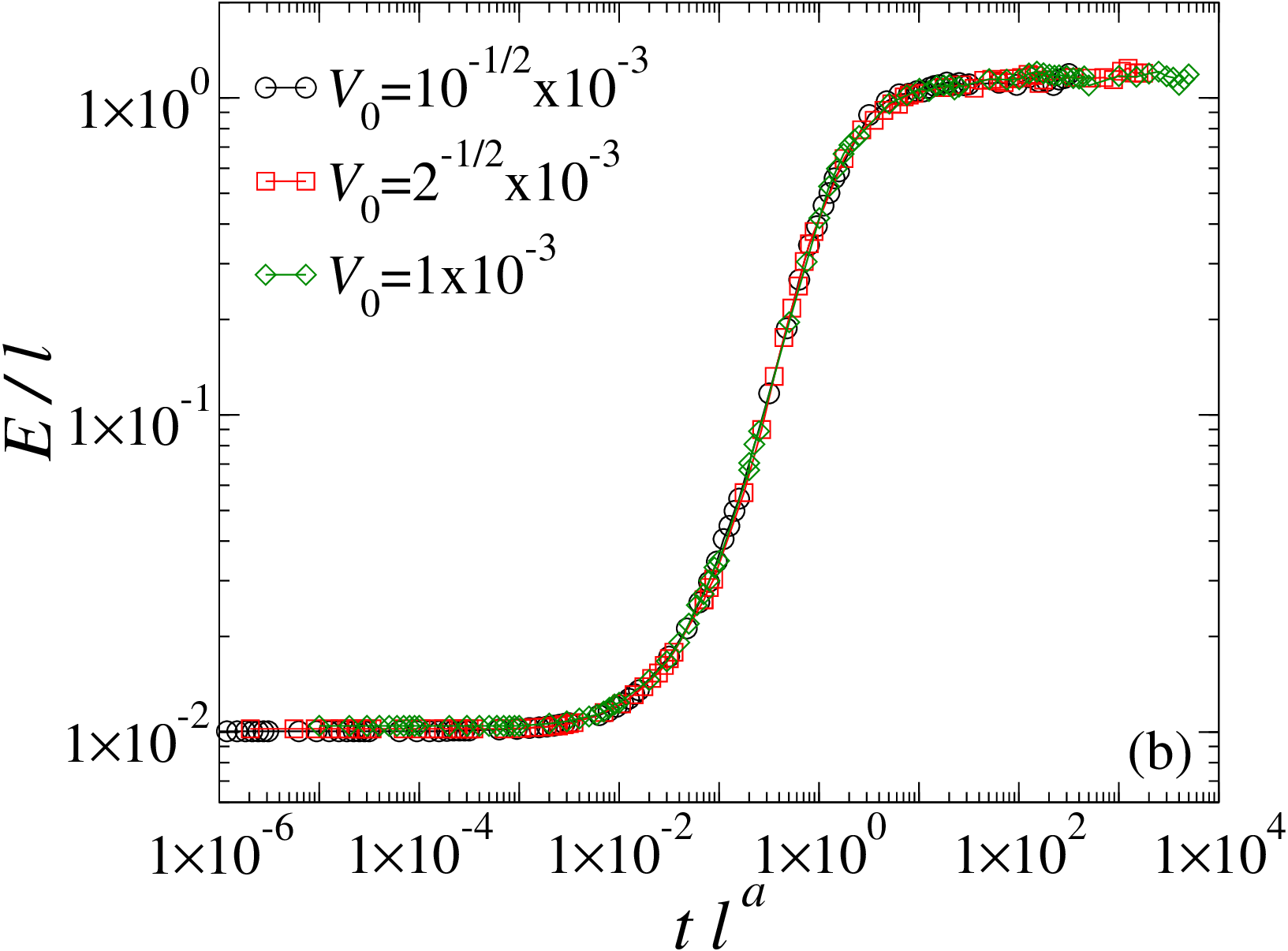}}
\caption{(Color online) The rescaled energy $E(t)/l$ as function of rescaled 
time $tl^a$ for three values of $\varepsilon$ in limit of long time. 
In (a) it is shown the collapse for $V_0<<\varepsilon$ ($V_0=1\times 10^{-6}$) 
and in (b) the collapse for three values of initial velocity $V_0>>\varepsilon$. 
We chose $l=\varepsilon^{-1/b}$ and the exponents are 
$a=1.49\pm 0.04$, $b=-0.99\pm 0.01$ and $c=-0.495\pm 0.005$. }
\label{fig2}
\end{figure}

Fig. \ref{fig2}(a) shows the rescaled energy $E(t,\varepsilon,V_0)/l$ as 
function of rescaled time $tl^a$. 
We can see that with these new coordinates the energy curves, 
originally depicted in Fig. \ref{fig1}(a), collapse onto 
a universal curve in the limit of long time, after an 
initial transient. 
We emphasize that the scaling behavior is valid only for small values of $\varepsilon$. 

The argument to obtain the $c$ exponent, namely $c=b/2$, is presented in the next section.
 Using this relation we obtain  that $c=-0.495\pm 0.005$. 
Fig. \ref{fig2}(b) shows the collapse of the curves depicted in 
Fig. \ref{fig1}(b) for the appropriate chosen initial velocities. 

Depending on initial velocity and phase, the velocity after a collision 
with the moving wall can be very small in such way that a direct collision 
occurs and gives an immediate increase in energy. 
This scenario is more common for small values of $V_0$ and $t$ and does not 
allow us to find a scaling description to the initial transient at small values of $t$. 
However, this transient is important because it affects the power-law growth of the energy 
between $t_1$ and $t_2$. 

\begin{figure}[t]
\centerline{\includegraphics[width=7cm,height=4.5cm]{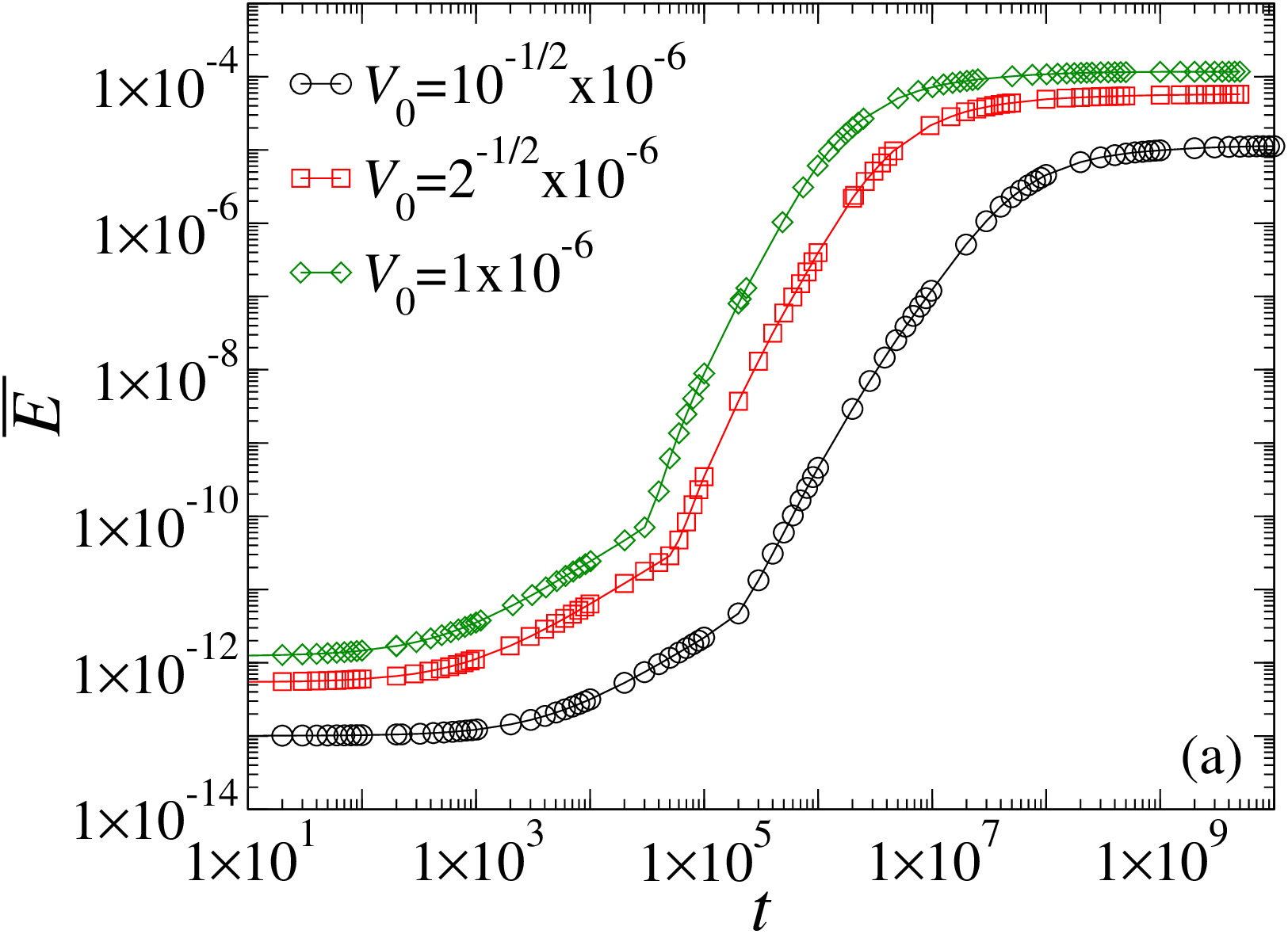}}
\vspace{0.4cm}
\centerline{\includegraphics[width=7cm,height=4.5cm]{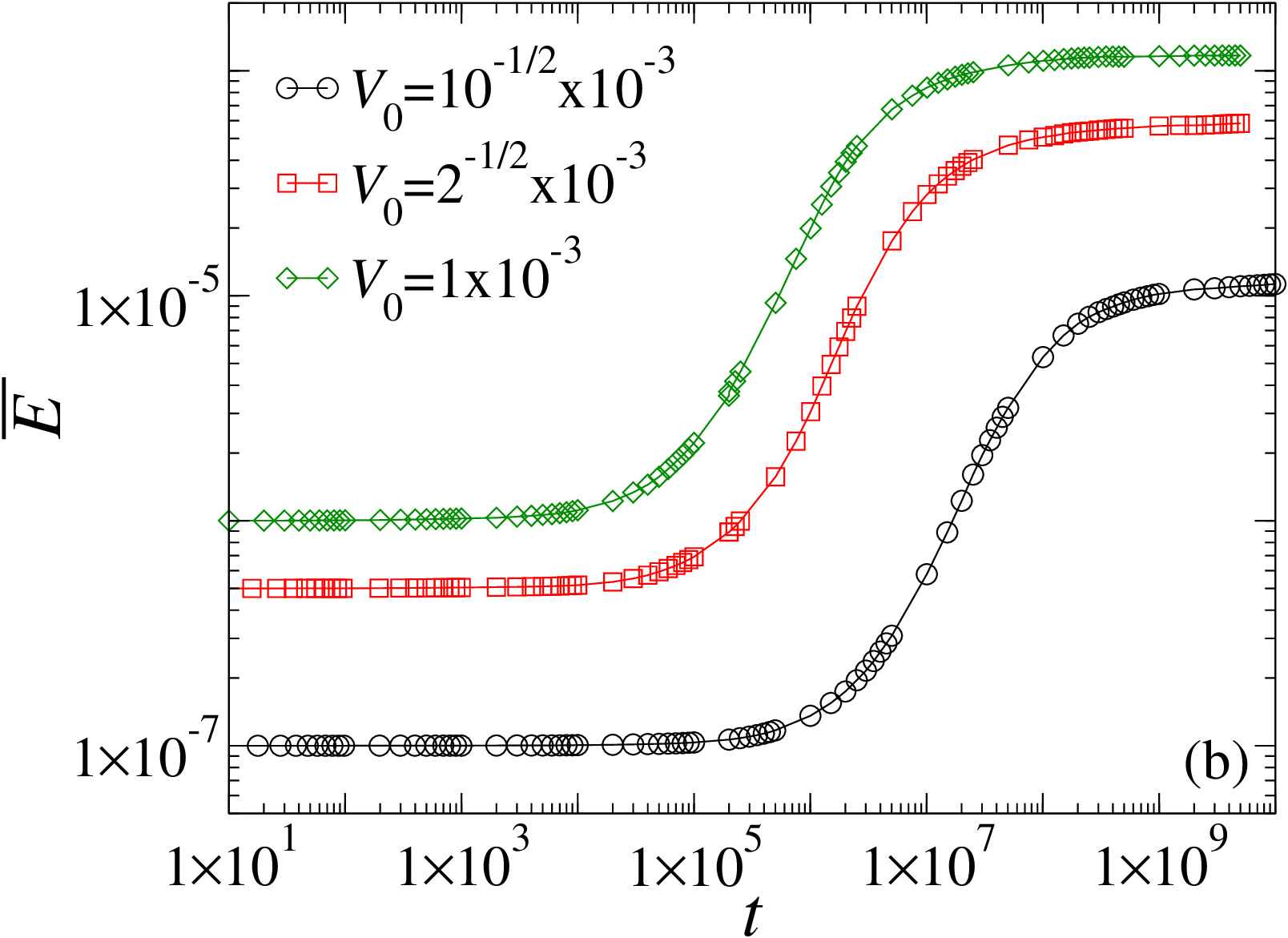}}
\caption{(Color online) The average energy $\overline{E}(t)$, defined in 
Eq. (\ref{eq5}), as function of time $t$ for three values of $\varepsilon$ and (a) 
three values of initial velocity $V_0<<\varepsilon$ and (b) three values of 
velocity $V_0>>\varepsilon$. The averages  were performed with 
$M=2\times 10^3$ samples. }
\label{fig3}
\end{figure}

\begin{figure}[t]
\centerline{\includegraphics[width=7cm,height=4.5cm]{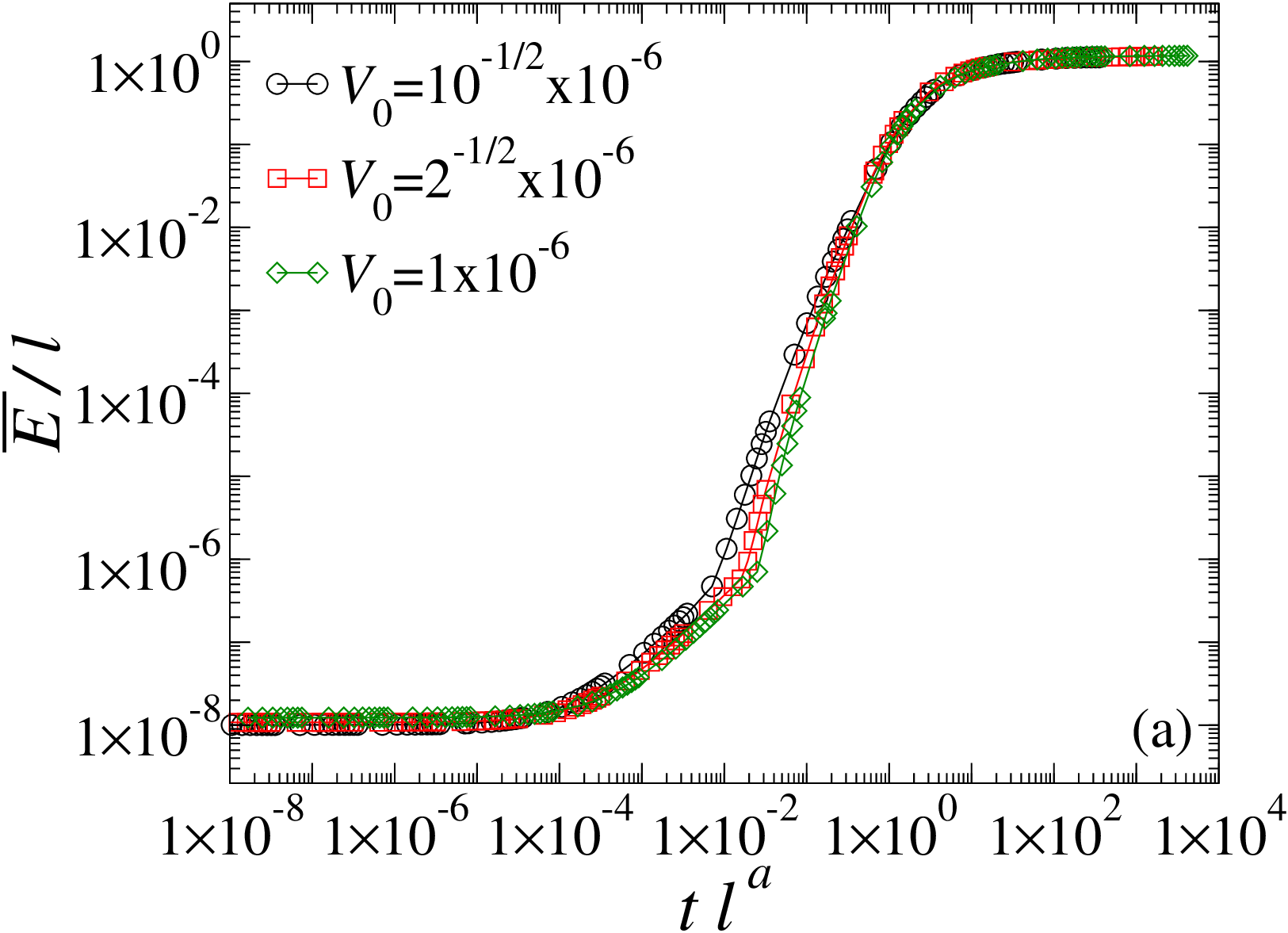}}
\vspace{0.4cm}
\centerline{\includegraphics[width=7cm,height=4.5cm]{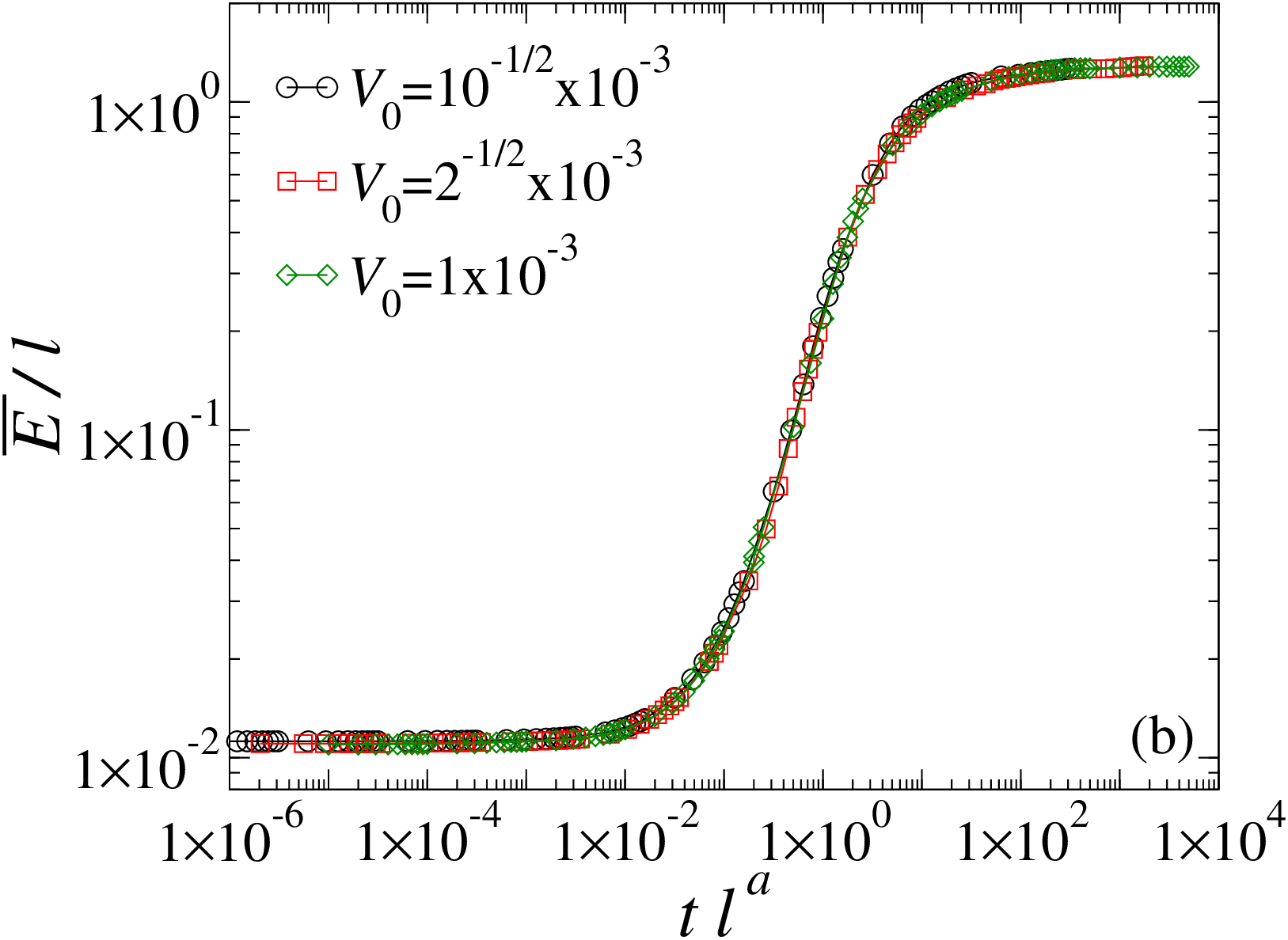}}
\caption{(Color online) Log-log plots of the rescaled energy $\overline{E}(t)/l$ 
as function of rescaled time $tl^a$ in limit of large time for three values 
of $\varepsilon$.  We have that (a) $V_0<<\varepsilon$ and (b) $V_0>>\varepsilon$. 
We chose $l=\varepsilon^{-1/b}$ and the exponents are 
$a=1.50\pm 0.04$, $b=-0.995\pm 0.003$ and $c=-0.498\pm 0.001$. }
\label{fig4}
\end{figure}

We can use Eqs. (\ref{eq6}) and (\ref{eq7}) to describe the average energy 
$\overline{E}(t,\varepsilon,V_0)$, defined in Eq. (\ref{eq5}), as a scaling 
function just changing $E$ by $\overline{E}$. 
Figs. \ref{fig3}(a) and \ref{fig3}(b) present the energy 
$\overline{E}(t,\varepsilon,V_0)$ as function of time $t$ 
for $V_0<<\varepsilon$ (small initial velocity) and $V_0>>\varepsilon$, respectively. 
The values of $\varepsilon$ are the same as those in Fig. \ref{fig1}(a). 
In Figs. \ref{fig4}(a) and \ref{fig4}(b) it is shown the rescaled energy 
as function of rescaled time in limit of large $t$ ($t>>t_1$). 
Following the same reasoning employed in analysis of the average energy 
$E(t,\varepsilon,V_0)$, we obtain the exponents 
$a=1.50\pm 0.04$, $b=-0.995\pm 0.003$ and $c=-0.498\pm 0.001$. 
Considering the uncertainties we observe that both averages 
$E(t,\varepsilon,V_0)$ and $\overline{E}(t,\varepsilon,V_0)$ are 
described by scaling functions with basically the same exponents set. 
Therefore, we will now use the average exponents 
$a=1.50\pm 0.04$, $b=-0.993\pm 0.007$ and $c=-0.497\pm 0.003$. 
Note that these values are different than the ones of 
the SFUM \cite{ladeira}, namely, 
$a_2=1.35\pm 0.05$, $b_2=-0.90\pm 0.03$ and $c_2=-0.45\pm 0.01$. 

\begin{figure}[t]
\centerline{\includegraphics[width=7cm,height=4.5cm]{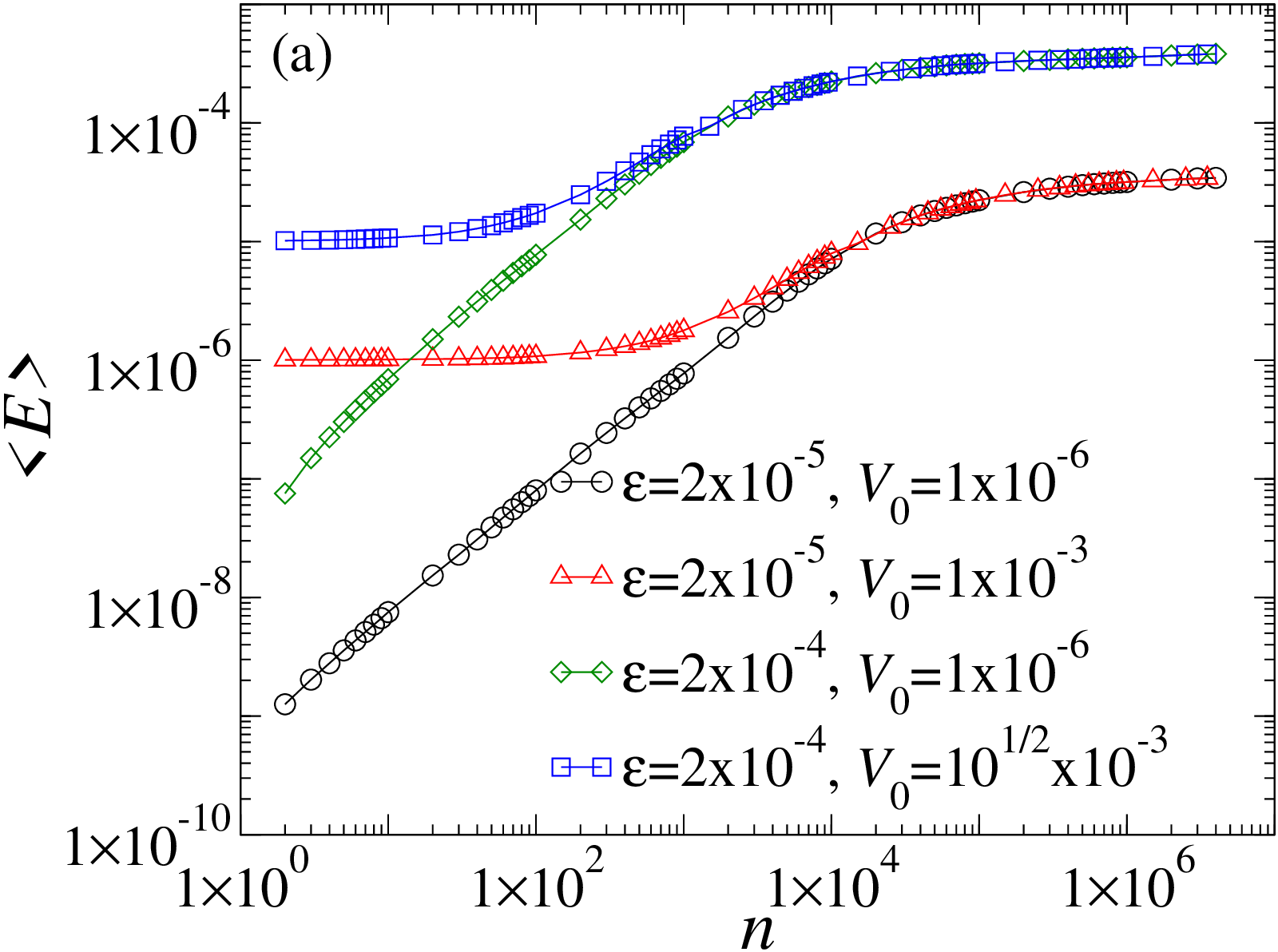}}
\vspace{0.4cm}
\centerline{\includegraphics[width=7cm,height=4.5cm]{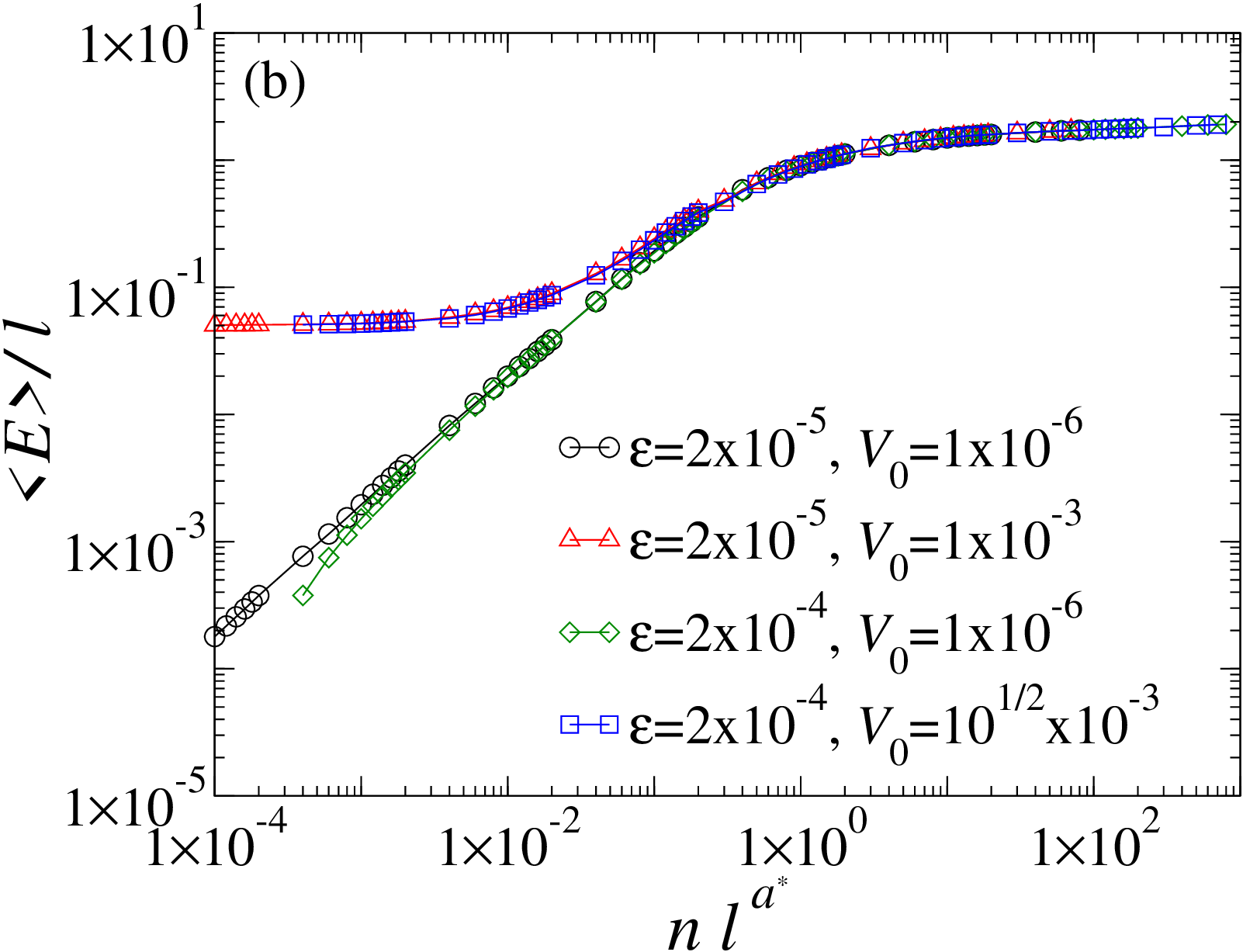}}
\caption{(Color online) (a) Average energy for two values of $\varepsilon$ and 
three values of initial velocity, $V_0$, as function of collision number, $n$. 
(b) The rescaled average energy as function of rescaled collision number shows 
that, after an initial transient, the energy curves collapse onto a universal curve.}
\label{fig5}
\end{figure}

Fig. \ref{fig5}(a) shows the average energy $<E>(n,\varepsilon,V_0)$ for two 
values of $\varepsilon$ and different initial velocities, including the situations 
(i) $V_0<<\varepsilon$ and (ii) $V_0>>\varepsilon$. 
 Now we have that
\begin{equation}
<E>(n,\varepsilon,V_0)=l<E>(l^{a^*}n,l^{b^*}\varepsilon,l^{c^*}V_0)~~.
\end{equation}
We can determine the exponents $a^*$ and $b^*$ from the energy curves for 
$V_0<<\varepsilon$. 
 When $V_0=1\times 10^{-6}$, we observe in Fig. \ref{fig5}(a) that 
 the average energy $<E>(n,\varepsilon,V_0)$ presents two regimes. For $n<n_x$ the 
energy has a power-law growth and for $n>>n_x$ the energy is constant. 
Moreover, we obtain that $n_x\propto\varepsilon^z$ with $z=-1.01\pm 0.02$. 

Since in the limit of large $n$ the energy depends only on $\varepsilon$, we have that 
$<E>(n,\varepsilon,V_0)=g(\varepsilon)\propto\varepsilon^\beta$ with $\beta=1.02\pm 0.02$. 
We can choose $l=\varepsilon^{-1/b^*}$ and rewrite the above equation as 
$<E>(n,\varepsilon,V_0)=\varepsilon^{-1/b^*}<E>(\varepsilon^{-a^*/b^*}n,1,\varepsilon^{-c^*/b^*}V_0)$. 
From this relation we obtain that $z=a^*/b^*=-1.01\pm 0.02$. In limit $n>>n_x$ we can write 
$<E>(n,\varepsilon,V_0)\propto\varepsilon^{-1/b^*}$. 
Therefore, we have that $\beta=-1/b^*=1.02\pm 0.02$. 
This implies that $b^*=-0.98\pm 0.02$ and $a^*=0.99\pm 0.04$. 
We use a connection between the simplified FUM and the standard map, 
described in the next section, 
to obtain the value of the exponent $c^*$, namely, $c^*=b^*/2=-0.49\pm 0.01$. 

When $V_0>>\varepsilon$ we observe in Fig. \ref{fig5}(a)  that  the energy curves 
present two characteristic iteration numbers, $n_x^\prime$ and $n_x^{\prime\prime}$. 
We can also observe that $n_x^{\prime\prime}\approx 0$ for $V_0<<\varepsilon$. 
Therefore, we must consider two situations: 
$n_x^{\prime\prime}<<n_x^\prime$, for small initial velocities ($V_0<<\varepsilon$), 
and $n_x^{\prime\prime}\sim n_x^\prime$, for $V_0>>\varepsilon$. 
The energy curves with $V_0=1\times 10^{-6}$ ($n_x^{\prime\prime}\approx 0$) 
present only two regimes: (1) a power-law growth for $n<<n_x^\prime$ and 
(2) a saturation regime for $n>>n_x^\prime$. 
On the other hand, for the energy curves with $V_0=1\times 10^{-3}$ and 
$V_0=10^{1/2}\times 10^{-3}$ ($n_x^{\prime\prime}< n_x^\prime$) 
shown in Fig. \ref{fig5}(a), we have three regimes: 
(1) the energy is basically constant for $n<<n_x^{\prime\prime}$, 
(2) for $n_x^{\prime\prime}<n<n_x^\prime$ the energy grows and begins to 
follow the curve of $V_0=1\times 10^{-6}$ and 
(3) the energy curves reach a saturation regime for $n>>n_x^\prime$. 

In Fig. \ref{fig5}(b) it is shown the rescaled energy $<E>/l$ as function of 
the rescaled interactions number $nl^{a^*}$. 
As we can observe, the energy curves, after a small initial transient, 
collapse onto a universal curve, even for 
$V_0>>\varepsilon$, with the exponents $a^*=0.99\pm 0.04$, $b^*=-0.98\pm 0.02$ and $c^*=-0.49\pm 0.01$. 
It is important to note that this set of exponents is the same, 
within the uncertainties, to that of the SFUM \cite{leonel2}. 
The exponents of the average energies as function of both time $t$ and collision number 
$n$ for the FUM and its simplification are shown in Table \ref{tab1}.

\begin{table}
\caption{Numerical estimations of the scaling exponents for both FUM and SFUM. 
The exponents $a$, $b$ and $c$ describe the scaling properties of the average energy 
as function of time $t$ while the scaling relations of the average energy as function of 
collision number $n$ are characterized by the exponents $a^*$, $b^*$ and $c^*$. 
The uncertainties are given between parenthesis. 
\label{tab1}}
\begin{center}
\begin{tabular}{|l|c|c|c|c|}
\hline

\multicolumn{1}{|c|} {} &  \multicolumn{2}{c|} {\bf FUM}   &  \multicolumn{2}{c|} {\bf SFUM}\\
   & $t$   & $n $  &  $t$ \cite{ladeira} & $n$ \cite{leonel2}\\ 
\hline 
$a$, $a^*$ &  $1.50  (4)$  & $ 0.99 (4)$    & $ 1.35 (5)$  &  $0.99  (3)$\\
$b$, $b^*$ & $-0.993 (7)$  & $-0.98 (2)$    & $-0.90 (3)$  & $-0.977 (6)$\\
$c$, $c^*$ & $-0.497 (3)$  & $-0.49 (1)$    & $-0.45 (1)$  & $-0.489 (3)$\\
\hline
\end{tabular}
\end{center}
\end{table}

\section{Analytical arguments}

Let us first derive a relation between the exponents $c$ ($c^*$)  and $b$ ($b^*$).
We follow the same lines of Leonel et al. \cite{leonel1}. 
Performing the variable change $I_n=2/V^*+2(V^*-V_n/{V^*}^2)$, where 
$V^*$ is a typical velocity near to the lowest spanning curve, and 
a linearization around $V^*$, the FUM transforms into a standard map which 
is described by $I_{n+1}=I_n - K_{\mathrm{eff}}\sin \phi_{n+1}$ and 
$\phi_{n+1}=\phi_n+I_n$. Here $K_{\mathrm{eff}}$ is an effective control parameter 
given by $K_{\mathrm{eff}}=4\varepsilon/{V^*}^2$. 
Note that the Standard model presents a transition from local to globally stochastic 
behavior at $K=K_{\mathrm{c}}\approx 0.972$. The values of $V^*$ in the 
lowest spanning curve of the FUM  furnish a $K_{\mathrm{eff}}$ with value approximately the same as $K_{\mathrm{c}}$, 
independent of $\varepsilon$. 
Then, we use the scaled variables $\varepsilon^\prime=l^{b}\varepsilon$ and 
${V^*}^\prime=l^{c}V^*$ to obtain 
$K_{\mathrm{eff}}=4\varepsilon^\prime/{{V^*}^\prime}^2=4(l^b\varepsilon)/(l^c V^*)^2$. 
This implies that $c=b/2$. 

Now we present an heuristic argument to support that $c^*=-1/2$. 
From Eq. (\ref{eq1}) we can write, for $n=1$, that 
\begin{equation}
V_1^2=V_0^2\pm 4\varepsilon V_0 \sin (\phi_1) + 4\varepsilon^2 \sin^2(\phi_1 )~~.
\label{jaff1}
\end{equation}
A similar equation can be obtained for $V_2^2$ by changing $V_0$ by $V_1$ and 
$\phi_1$ by $\phi_2$. Then, we replace $V_1^2$ by Eq. (\ref{jaff1}). This iteration procedure can 
be done for $V_n^2$ with arbitrary $n$. If now we take the average in the ensemble of initial 
phases $<V_n^2>$, we always find a sum of 
three kinds of terms: (i) $V_0^2$, (ii) a set of terms  $\pm V_0 \varepsilon<\sin(\phi_j)>$ 
and (iii) a set of terms  $\pm\varepsilon^2<\sin(\phi_i)\sin(\phi_j)>$. Observe that 
$<\sin(\phi_j)>$ and $<\sin(\phi_i)\sin(\phi_j)>$ take values in the interval $[-1,1]$. 
Since we are in the region below the first spanning curve, the maximal initial value 
$V_{0,\mathrm{max}}$ must be of order of $V^*\approx\varepsilon^{1/2}$. 
Therefore, for small $\varepsilon$, $V_0\approx V_{0,\mathrm{max}}>>\varepsilon$ and $n$ small 
enough we have that $V_0^2\lesssim ~ <V_n^2> ~ \gtrsim V_0^2$, implying that  $<V_n^2>\approx V_0^2$. 
 Assuming that the scaling relation is valid in 
this limit, we obtain the relation $<V_n^2>\approx l^{1+2c^*}<V_n^2>$, which furnishes 
$c^*=-1/2$. It is worth mentioning that the numerical results for $V_0>>\varepsilon$ 
shows that the scaling occurs in the beginning of the simulation without any transient 
(see Fig. (\ref{fig5}). Moreover, using the  result $c^*=b^*/2$, we also obtain that 
$b^*=-1$. 

Finally we present an argument that relates the exponents $a^*$ and $a$, the 
scaling dimensions of variables $n$ and $t$, respectively. 
As the average energy presents a saturation regime at large values of time, 
we can define the average velocity 
$\overline{V}=\sqrt{\overline{E}}$ and $\Delta t$ 
as the average time between two arbitrary collisions. We can also define 
$\overline{L}$ as the average 
distance that the particle travels between these collisions in such way that we can 
write $\overline{V}\Delta t/\overline{L}=\Delta n$, where $\Delta n$ is the average 
collisions number with the moving wall that take place in the time interval $\Delta t$. 
In terms of the rescaled variables we have that 
$\overline{V}^\prime\Delta t^\prime/\overline{L}^\prime=\Delta n^\prime$, or
$l^c\overline{V} l^a\Delta t/\overline{L}=l^{a^*} \Delta n$. 
As $\overline{L}^\prime=\overline{L}$ we have, therefore, that the exponent 
of collision number, $a^*$, is related to the exponent of time, $a$, by $a^*=a+c$. 
This result is in good agreement with the simulations. 

\section{Conclusions}

We studied the scaling properties of the chaotic sea below the lowest energy 
spanning curve of FUM considering average energies as function of time 
$t$ and collision number $n$. 
In limit of large $t$ ($t>>t_1$), the average energies $E$ and $\overline{E}$ of FUM can be described by 
scaling functions with exponents 
$a\approx 3/2$, $b\approx -1$ and $c\approx -1/2$ (see Table \ref{tab1}). 
This values are different than the ones of the SFUM \cite{ladeira}, namely, 
$a_2=1.35\pm 0.05$, $b_2=-0.90\pm 0.03$ and $c_2=-0.45\pm 0.01$. 
The scaling descriptions of the average energies are also hold on variable $n$, with exponents 
$a^*\approx 1$, $b^*\approx -1$ and $c^*\approx -1/2$. 
These values are basically the same as those of the SFUM \cite{leonel2}. 
We employed some analytical arguments  to determine 
the exponents $c^*=-1/2$ and $b^*=-1$. 
We observe also that, in the full model, the scaling exponents related 
to variables $\epsilon$ and $V_0$ are, within the uncertainties, the same 
for both $n$ and $t$ analyses, or $b^*\approx b$ and $c^*\approx c$. 
We can also note that the exponents related to the variables $t$ and $n$ 
($a$ and $a^*$, respectively) are note the same. 
However, we provide an heuristic argument that 
establishes a connection between these 
exponents by the relation $a^*=a+c$.
It is important to stress that the scaling analysis only holds for small 
values of the rescaled amplitude $\varepsilon$, close enough to the 
integrable ($\varepsilon=0$) to non-integrable  ($\varepsilon\neq 0$) transition. 
Moreover, we observe that the energy of the FUM 
is constant up to a time $t_1$, grows to a time $t_2$ and, then, reaches a stationary value. 
In the SFUM the average energy, differently of FUM, presents a slow decay for large 
$t$ ($t>>t_2$) \cite{ladeira}. 
This striking result is a consequence of the 
approximation used to define the SFUM. As the oscillating wall is considered fixed in space,  
the time between collisions is $2/V$ and successive 
collisions do not occur. 
Therefore, eventually after a collision with the moving 
wall, the particle has very low velocity $V$ and remains for a long time ($2/V$) with low energy, 
originating a slow decay in energy for $t>>t_2$. 
In the full model it is different: if after a collision with the moving wall the particle 
losses almost all its energy, then a successive collision occurs increasing the energy of 
the particle and the decay in energy at $t>>t_2$ is not observed.  
The analyses on variable $n$ furnish basically the same results for 
both FUM and SFUM because 
between one arbitrary collision and the next one we always have that 
$\Delta n=1$, independently if the collisions are direct (successive) or indirect. 

Summarizing, 
we employed scaling analyses to describe the properties of average energies 
as function of $t$ and as function of $n$ in 
regime of small amplitudes of oscillation of 
the moving wall, $\varepsilon\approx 0$, for the full Fermi-Ulam model. 
We observe that 
i) the scaling exponents related to the variables $\varepsilon$ 
and $V_0$ are basically the same for both $t$ and $n$ analyses, namely, 
$b\approx b^*\approx -1$ and $c\approx c^*\approx -1/2$. 
ii) We also observe that the scaling exponents related to the variables $t$ and $n$ 
are not the same, $a\approx 3/2$ and $a^*\approx 1$, respectively, and, by 
a simple analytical reasoning, we show that these exponents 
are connected by the relation $a^*=a+c$. 
iii) Performing some analytical analyses on variable $n$ we 
derive that $b^*=-1$ and $c^*=-1/2$. 
iv) Considering also the results of the simplified Fermi-Ulam model, 
\cite{leonel2} and \cite{ladeira}, we observe that the scaling analyses of FUM and 
SFUM on variable $t$ are characterized by different exponents sets, 
while the analyses on variable $n$ furnish, basically, the same 
exponents set for both FUM and SFUM. 
v) We observe also that the successive collisions play an important 
rule when the Fermi-Ulam model is studied on variable $t$ by, mainly, preventing the 
energy decay for long times. 
vi) In the analysis on variable $n$ direct (successive) and indirect collisions play a 
similar rule and, therefore, the energy curves as function of $n$ present, for both 
FUM and SFUM, the same behavior and they are described by a single set of scaling exponents. 

We thank to J.A. Plascak for the careful reading of the manuscript. 
D.G.L. and J.K.L.S. were partially supported by Conselho Nacional de Pesquisa (CNPq).
J.K.L.S. also thanks to Funda\c c\~ao de Amparo \`a Pesquisa do Estado de Minas Gerais (FAPEMIG).\\
$^{\ast}$      Electronic address: dgl@fisica.ufmg.br\\
$^{\dagger}$   Electronic address: jaff@fisica.ufmg.br\\

\end{document}